\documentclass[a4paper,fleqn]{cas-dc}

\usepackage[authoryear,longnamesfirst]{natbib}

\usepackage[utf8]{inputenc}
\usepackage{textcomp}

\usepackage{graphicx}
\usepackage{caption}
\usepackage{subcaption}
\usepackage{savesym}
\usepackage{amsmath}
\savesymbol{Bbbk}
\usepackage{amssymb}
\restoresymbol{SYM}{Bbbk}
\usepackage[version=4]{mhchem}
\usepackage{siunitx}
\usepackage{longtable,tabularx}
\usepackage{bm}
\usepackage{color}
\usepackage{dsfont}
\usepackage{multicol}

\usepackage[english]{babel}
\addto\captionsenglish{}
\usepackage{theoremref}
\usepackage{hyperref}

\newtheorem{thm}{Theorem}
\newtheorem{lem}{Lemma}

\newtheorem{prob}{Problem}

\newtheorem{asm}{Assumption}
\newtheorem{rem}{Remark}

\def\tsc#1{\csdef{#1}{\textsc{\lowercase{#1}}\xspace}}
\tsc{WGM}
\tsc{QE}

\begin{document}
\let\WriteBookmarks\relax
\def\floatpagepagefraction{1}
\def\textpagefraction{.001}

\shorttitle{3D Path Following and L1 Adaptive Control for Underwater Vehicles}    

\shortauthors{3D Path Following and L1 Adaptive Control for Underwater Vehicles}  

\title [mode = title]{3D Path Following and L1 Adaptive Control for Underwater Vehicles}  

\tnotemark[1] 

\tnotetext[1]{This research was supported by the Office of Naval Research, grants N000141912106 and N000142112091.} 

%

\author[1]{Nicholas Rober}



\ead{nrober@mit.edu}



\affiliation[1]{organization={University of Iowa},
            addressline={103 S. Capitol Street}, 
            city={Iowa City},
            postcode={52242}, 
            state={Iowa},
            country={United States}}

\author[1]{Maxwell Hammond}


\ead{maxwell-hammond@uiowa.edu}






\author[1]{Venanzio Cichella}


\ead{venanzio-cichella@uiowa.edu}

\ead[url]{venanziocichella.com}


\author[1]{Juan E. Martin}


\ead{juan-martin@uiowa.edu}



\author[1]{Pablo Carrica}


\ead{pablo-carrica@uiowa.edu}


\begin{abstract}
This paper addresses the problem of guidance and control of underwater vehicles. A multi-level control strategy is used to determine (1) outer-loop path-following commands and (2) inner-loop actuation commands. Specifically, a line-of-sight path-following algorithm is used to guide the vehicle along a three-dimensional path, and an $\mathcal{L}_1$ adaptive control algorithm is used to determine the low-level rudder commands to accomplish path following. The performance bounds of these outer- and inner-loop control algorithms are presented. Numerical results obtained using a physics-based Simulink model are used to aid in visualization of the control algorithm's performance.
\end{abstract}




\maketitle

\section{Introduction}
The increasingly complex technological capabilities of autonomous underwater vehicles (AUVs) have allowed them to become important tools for various types of missions in aquatic and marine environments. Namely, AUVs provide new opportunities in applications including bathymetry (\cite{ma2018auv,caress2008high,henthorn2006high}), archaeology (\cite{tsiogkas2014efficient,bingham2010robotic}), oceanographic exploration (\cite{kunz2008deep,mcphail2009exploring}), underwater maintenance and infrastructure monitoring (\cite{palomer2019inspection,kim2013real}), and defense (\cite{williams2010optimal,djapic2010using,munafo2017auv}). In order to successfully accomplish these missions, it is often important for the AUV to operate in conditions that are inherently hazardous to the vehicle. This provides an interesting problem from a control perspective because it requires the design of a controller with known performance bounds that can be used to avoid collisions with obstacles and interaction with the surface.

In order for AUVs to complete mission objectives, it is important that they are capable of following desired spatial trajectories that avoid obstacles and navigate them to specified points of interest. Known strategies used for the motion control of AUVs include waypoint tracking, trajectory tracking, and path following. \emph{Waypoint tracking} (see, for example, \cite{wang2018switching,rout2016narmax}) achieves motion control by directing the AUV through a series of target points distributed throughout the environment. \emph{Trajectory tracking} (e.g., \cite{elmokadem2017terminal,shen2017trajectory,guerrero2019trajectory}) allows the vehicle to more precisely track a 3D trajectory parameterized by time. Finally, \emph{path following} (see \cite{abdurahman2017switching,peng2017output,paliotta2018trajectory,encarnacao20003d,lapierre2003nonlinear}), which is used in this work, is similar to trajectory tracking, but gives an additional degree of freedom by introducing a path-following parameter that enables the vehicle to track a virtual target on the path. The progression of the virtual target can be determined on-line via a control law, thus allowing the vehicle to track a reference which is dependant on its position and velocity, as opposed to a reference which simply progresses with mission time, as is the case with trajectory tracking.

The guidance strategy used in this paper is a path-following technique borrowing from work initially done in \cite{micaelli1993trajectory}, which solves the path-following problem for wheeled robots. This is done by attaching a Frenet-Serret frame to a target point, referred to as the \emph{virtual target}, which moves along the path and provides a desired position and orientation. This idea was then expanded to 3D and implemented on unmanned aerial (see \cite{kaminer1998trajectory}) and underwater vehicles (see \cite{silvestre2000multi}). The method of placing the virtual target along the path was then improved upon by \cite{lapierre2006nonsingular}, which uses the vehicle's linear and angular velocities to determine the rate of progression of the virtual target and reduce the path-following error to zero. Finally, in \cite{lee2010geometric}, the authors use a coordinate-free approach to avoid the singularity problems associated with using local vehicle coordinates. Techniques from these previous works were used to develop the path-following algorithm outlined in \cite{cichella2011geometric,kaminer2017time,cichella20113d,cichella20133dmultirotor}, which constitute the high-level motion-control algorithm used in here.

While control strategies for underwater vehicles range from direct position tracking, see \cite{silvestre2007depth}, to multi-layer path-following methodologies (\cite{rober2021three}), each strategy employs a vehicle autopilot which is tasked with controlling vehicle dynamics to track a desired reference. Some of the early work on autopilot design for underwater vehicles used sliding-mode control methods to confine the vehicle's behavior to a set of sliding surfaces with desired properties (see, for example, \cite{slotine1145131,cristi1990adaptive,healey1993multivariable}). Sliding-mode control is still a common option for autopilot design and recent studies have sought to reduce the chattering phenomenon, which is a common problem encountered when employing this control strategy, see \cite{cui2016adaptive,wang2016multivariable,bessa2008depth}. Proportional-integral-derivative (PID) control strategies have also been used for tracking control, and while their performance can be inferior to other strategies and they take time to properly tune, they are relatively simple to implement (\cite{jalving1994ndre,martin2017nonlinear}). Additionally, with the development of artificial intelligence and its increasing prevalence in controls-related applications, learning-based algorithms have been the subject of recent investigation by, for example, \cite{aras2015depth,cui2017adaptive,wu2018depth}.

The strategy used to develop the autopilot controller in this paper is based on the $\mathcal{L}_1$ adaptive control architecture presented in \cite{hovakimyan2010}. $\mathcal{L}_1$ control decouples robustness and adaptation, thus allowing for fast and robust adaptation in the presence of system uncertainty and external disturbances. Moreover, while $\mathcal{L}_1$ control schemes can be used directly to control a system (see \cite{gregory2009l1}), they can also be used to augment a system with an existing autopilot to improve its performance, such as in \cite{kaminer2010path}. This control-augmentation strategy can be especially helpful when considering that many commercially available systems have built-in autopilots that are not easily modified (\cite{girard2007autopilots}). In these cases, $\mathcal{L}_1$ control can be used to augment the existing autopilot by feeding it an adaptive reference that is used to give the system desired properties. 

Additionally, the ability to augment an existing controller to improve its performance comes with a relatively low cost during the design phase of the control algorithm: once the desired properties of the closed-loop system are determined, there is typically very little tuning required for $\mathcal{L}_1$ control design. Finally, the robustness properties of the $\mathcal{L}_1$ control formulation, along the fact that its boundedness properties can be used in conjunction with those of the path-following controller to provide an overall bound on the performance of the combined controller, make the $\mathcal{L}_1$ controller an excellent choice for inner-loop control of a path-following strategy. Thus the inner-loop controller formulated in this paper uses $\mathcal{L}_1$ control to augment an autopilot and allow the generic underwater vehicle model Joubert BB2 introduced in \cite{carrica2019} to track reference steering commands from a path-following controller.

This paper extends previous work by applying the sampled-data $\mathcal{L}_1$ adaptive control structure introduced in \cite{jafarnejadsani2018robust} in conjunction with the path-following guidance algorithm proposed in \cite{rober2021three}. We present the stability and performance bounds for each control algorithm, which can be combined to give an overall performance bound for the vehicle's motion controller as a whole. Finally, we demonstrate the control algorithm  and its improvements over a commercial autopilot by presenting numerical results  with and without adaptation employing a recently-developed physics-based Simulink model introduced in \cite{kim2021development}. 

The organization of this paper is as follows. In Section \ref{sec:pf_problem} we formulate and solve the problems associated with path-following. Section \ref{sec:il_problem} then introduces a sampled-data adaptive controller and provides analysis of its performance. Next, Section \ref{sec:results} demonstrates the controller performance through a series of numerical simulations which highlight how the control strategy performs in a variety of simulations. Finally, Section \ref{sec:conclusion} gives concluding remarks and outlines goals for future work.

\section{Path-Following Problem}\label{sec:pf_problem}

\begin{figure}
    \centering
    \includegraphics[width=0.47\textwidth]{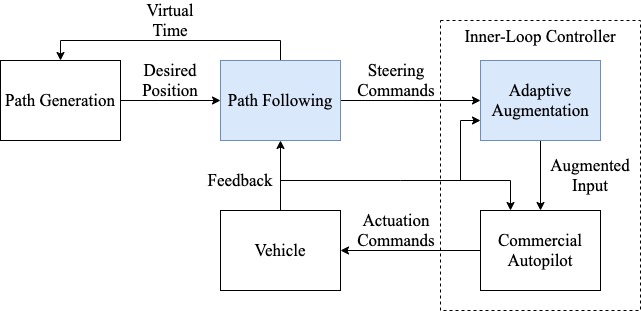}
    \caption{The overall control structure highlighting the path-following controller and the adaptive augmentation algorithm}
    \label{fig:pf_block_diagram}
\end{figure}

Fig. \ref{fig:pf_block_diagram} first shows a high-level visualization of the overall control strategy, highlighting the two components addressed in this paper: the path-following algorithm and the adaptive control scheme.
In this section we outline the formulation for the path-following problem and subsequently provide the solution that gives steering commands for the vehicle to converge to and follow a generic 3D path. As a preliminary to the introduction of the path-following algorithm, we define the \emph{hat map}~$(\cdot)^{\wedge}:\mathbb{R}^3\rightarrow\text{so}(3)$ as
$$
(\bm{\xi})^{\wedge} = \left[\begin{array}{ccc} 0 & -\xi_3 & \xi_2 \\ \xi_3 & 0 & -\xi_1 \\ -\xi_2 & \xi_1 & 0 \end{array}\right]
$$
for $\bm{\xi}=[\xi_1,~\xi_2,~\xi_3]^{\top}\in\mathbb{R}^3$, where $\text{so}(3)$ defines the set of skew symmetric matrices defined over $\mathbb{R}$. The \emph{vee map}~$(\cdot)^{\vee}:\text{so}(3)\to\mathbb{R}^3$ is the inverse of the hat map. A property of the hat and vee maps used in this paper is given below:
\begin{equation} 
\begin{split}
\text{tr}\left[{(\bm{\xi})^{\wedge}\bm{T}}\right]
& = \text{tr}\left[{\bm{T}(\bm{\xi})^{\wedge}}\right]
 = \frac{1}{2}\text{tr}\left[{(\bm{\xi})^{\wedge}(\bm{T}-\bm{T}^{\top})}\right] \\ & 
= -\bm{\xi}\cdot\left(\bm{T}-\bm{T}^{\top}\right)^{\vee}\,\label{eq:hatvee.prop1}
\end{split}
\end{equation}
for any $\bm{\xi}\in\mathbb{R}^3$, and $\bm{T}\in\mathbb{R}^{3\times3}$, where $\text{tr}(\cdot)$ is the trace of a matrix (see \cite{cichella2011geometric}). 

\subsection{Path-Following Problem Formulation}
\label{sec:pf_formulation}
\begin{figure}
    \centering
    \includegraphics[width=0.47\textwidth]{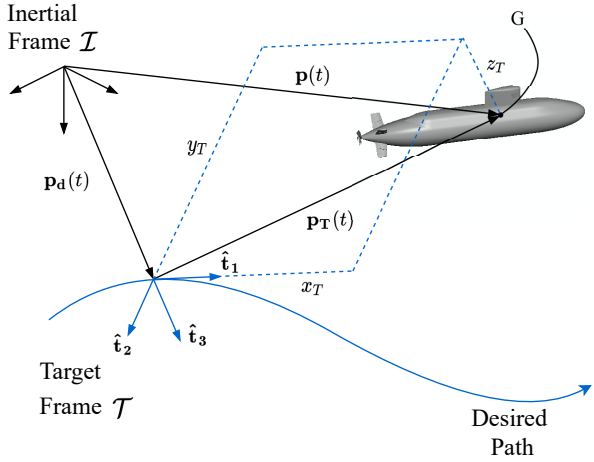}
    \caption{Geometry associated with the path-following problem, borrowed from \cite{rober2021three}}
    \label{fig:vector_summary}
\end{figure}

First, Fig. \ref{fig:vector_summary} depicts the geometry of the path-following problem, showing important vectors used to determine the path-following error.
We first introduce $\mathcal{I}$ as the inertial reference frame and let $\bm{p_d}:[0,T_f] \to \mathbb{R}^3$ be the desired \emph{geometric path} defined in this frame. To enable path following, $\bm{p_d}$ is parametrized by the variable $\gamma:\mathds{R}^+\rightarrow[0,T_f]$ which is referred to as \emph{virtual time} and allows the control law to shift the desired point on the path, referred to as the \emph{virtual target}, as a function of the vehicle's state. Thus, for a given value of $\gamma(t)$, the virtual target is represented by $\bm{p_d}(\gamma(t))$. Hereafter, we omit the use of ``$(t)$'' except where it is necessary for clarity. Next we construct a parallel transport frame (see \cite{kaminer2017time} and references therein) which defines the orientation of the virtual target, namely the \emph{target frame} $\mathcal{T}$. The $\mathcal{T}$ frame has its origin at $\bm{p_d}(\gamma)$ and its orientation with respect to the $\mathcal{I}$ frame is represented by the rotation matrix $\bm{R}_T^I(\gamma) \triangleq [\bm{\hat t}_1(\gamma),\bm{\hat t}_2(\gamma),\bm{\hat t}_3(\gamma)]$, where $\bm{\hat t}_1(\cdot)$ is the unit vector tangent to the path and thus aligned with the velocity of the virtual target, i.e., $\bm{\hat t}_1(\gamma) = \frac{\bm{p_d}^{\prime}(\gamma)}{\Vert \bm{p_d}^{\prime}(\gamma) \Vert}$, with $\bm{p_d}^{\prime}(\gamma) = \frac{d\bm{p_d}(\gamma)}{d \gamma}$. To construct $\bm{R}_T^I(\gamma)$, the unit vectors $\bm{\hat t}_2(\gamma)$ and $\bm{\hat t}_3(\gamma)$ are made to be orthonormal to $\bm{\hat t}_1 (\gamma) $ and are found using the following relationships:
\begin{equation*} 
\begin{split} 
\bm{\hat t}_1^\prime(\gamma) & = k_1(\gamma) \bm{\hat t}_2(\gamma) + k_2(\gamma) \bm{\hat t}_3(\gamma) \\
\bm{\hat t}_2^\prime(\gamma) & = -k_1(\gamma) \bm{\hat t}_1(\gamma) \\
\bm{\hat t}_3^\prime(\gamma) & = -k_2(\gamma) \bm{\hat t}_1(\gamma) \\
\kappa (\gamma) & =  \left( k_1^2(\gamma) + k_2^2(\gamma) \right)^{\frac{1}{2}} \\
\tau (\gamma) & = -\frac{d}{d \gamma} \left( \tan^{-1} \left( \frac{k_2 (\gamma)}{k_1 (\gamma)} \right) \right)
\end{split} \, 
\end{equation*}
where $\kappa(\gamma)$ and $\tau(\gamma)$ are the curvature and torsion of the path at $\gamma$. The angular velocity of $\mathcal{T}$ with respect to $\mathcal{I}$ resolved in $\mathcal{T}$, i.e., $\bm{\omega}_T$ is, computed using the above relationship as 
\begin{equation} \label{eq:omegaT} 
{\bm{\omega}_T} = [0, -k_2(\gamma) \dot{\gamma}, k_1(\gamma) \dot{\gamma}]^\top \,
\end{equation}
The geometric path can be constrained to satisfy the speed and angular rate limits given by 
\begin{equation} \label{eq:boundsTT} 
\begin{split} 
& 0 < v_{T,\min} \leq \Vert \bm{p_d}^\prime (\gamma) \Vert \leq v_{T,\max} \, \\ \, \quad & \Vert {\bm{\omega}_T}(\gamma) \Vert = \sqrt{k_1^2(\gamma)+k_2^2(\gamma)} \leq \omega_{T,\max}
\end{split} 
\end{equation}
and can be generated off-line by a path-generation algorithm as is done in \cite{cichella2019optimal}. In contrast, the position of the virtual target can be determined on-line by controlling the derivative of virtual time $\dot{\gamma}$ as will be discussed later. This allows the vehicle to target a point on the path that is dependent on the vehicle's position and velocity relative to the path.

Next we introduce the \emph{flow} frame $\mathcal{W}$ which has its origin $\bm{p} \triangleq [x_W,~y_W,~z_W]: \mathbb{R}^+\to\mathbb{R}^3$ at the center of mass of the vehicle. Its orientation is given by $\bm{R}_W^I \triangleq [\bm{\hat w}_1,\bm{\hat w}_2,\bm{\hat w}_3]$ and is configured such that its x-axis is aligned with the vehicle's velocity, i.e., $\bm{v}_W = [v,0,0]^\top$, where $v \in \mathbb{R}^{+}$ is the vehicle's speed. The angular velocity of $\mathcal{W}$ with respect to $\mathcal{I}$ resolved in $\mathcal{W}$ is then denoted by $\bm{\omega}_W \triangleq [p,q,r]^{\top}$.

Having developed reference frames corresponding to the virtual target and the vehicle, we can now introduce the path-following position error, resolved in the $\mathcal{T}$ frame, as:
\begin{equation} \label{eq:positionerror}
\bm{p_T} = \bm{R}^{T}_{I}(\bm{p}-\bm{p_d}(\gamma)) \triangleq [x_T,y_T,z_T]^\top
\end{equation}
From the dynamics of the virtual target and the vehicle, the position error dynamics can be derived as
\begin{equation} \label{eq:positionerrordynamics}
\begin{split}
    \dot{\bm{p_T}}  & =
    \dot{\bm{R}}_{I}^{T}(\bm{p}-\bm{p_d}(\gamma)) + {\bm{R}}_{I}^{T} \dot{\bm{p}} - {\bm{R}}_{I}^{T} \dot{\bm{p}}_{\bm{d}} 
    \\ & = -{\bm{\omega}}_T \times \bm{p_T} + {\bm{R}}_{W}^{T} \begin{bmatrix} v \\ 0 \\ 0 \end{bmatrix} - 
    \begin{bmatrix} \Vert \bm{p_d}^\prime(\gamma) \Vert \dot{\gamma} \\ 0 \\ 0 \end{bmatrix}
 \end{split}
\end{equation}
Next, to complete the formulation of the path-following problem, we derive the path-following attitude error. This is done by introducing a desired frame $\mathcal{D}$ attached to the vehicle's center of mass and specifies the desired orientation of the vehicle's flow frame. The orientation of $\mathcal{D}$ is represented by the rotation matrix from $\mathcal{D}$ to $\mathcal{I}$ $\bm{R}_D^I \triangleq [\bm{\hat b}_{1D},\bm{\hat b}_{2D},\bm{\hat b}_{3D}]$ with the basis vectors given by
\begin{align*} 
{\bm{\hat b}_{1D}} \triangleq \frac{d\bm{\hat t}_{1}-y_T\bm{\hat t}_{2}-z_T\bm{\hat t}_{3}}
{(d^2+y_T^2+z_T^2)^\frac{1}{2}} &&
{\bm{\hat b}_{2D}} \triangleq \frac{y_T\bm{\hat t}_{1}+d\bm{\hat t}_{2}}
{(d^2+y_T^2)^\frac{1}{2}} 
\end{align*}
and ${\bm{\hat b}_{3D}} = {\bm{\hat b}_{1D}} \times \bm{\hat b}_{2D}$. The \emph{characteristic distance} $d>0$ is a constant design parameter which influences the approach behavior of the path-following algorithm. As will be discussed in depth later, one of the objectives of the path-following controller is to align the vehicle's velocity with the desired direction by aligning $\bm{\hat w}_1$ with $\bm{\hat b}_{1D}$. Thus, as $d$ is increased, $\bm{\hat b}_{1D}$ is directed to align more closely with $\bm{\hat t}_{1}$. Conversely, as $d$ is decreased, the $y_T$ and $z_T$ error components, collectively referred to as the \emph{cross-track error}, have a larger influence on the desired direction, causing the vehicle to more aggressively approach the path. 

The orientation of $\mathcal{D}$ with respect to $\mathcal{T}$ can then be constructed from the definitions $\bm{\hat b}_{1D}$, $\bm{\hat b}_{2D}$ and $\bm{\hat b}_{3D}$ and is written as
\begin{equation*}
	\bm{R}_{D}^{T} = 
	\begin{bmatrix}
		\frac{d}{(d^2+y_T^2+z_T^2)^\frac{1}{2}} & \frac{y_T}{(d^2+y_T^2)^\frac{1}{2}} & \frac{dz_T}						{(d^2+y_T^2+z_T^2)^\frac{1}{2}(d^2+y_T^2)^\frac{1}{2}} \\
		\frac{-y_T}{(d^2+y_T^2+z_T^2)^\frac{1}{2}} & \frac{d}{(d^2+y_T^2)^\frac{1}{2}} & \frac{-y_Tz_T}						{(d^2+y_T^2+z_T^2)^\frac{1}{2}(d^2+y_T^2)^\frac{1}{2}} \\
		\frac{-z_T}{(d^2+y_T^2+z_T^2)^\frac{1}{2}} & 0 & \frac{(d^2+y_T^2)^\frac{1}{2}}{(d^2+y_T^2+z_T^2)^\frac{1}{2}}
	\end{bmatrix} \, 
\end{equation*}
Notice that while $\bm{R}_D^T$ does not contain $x_T$, this error term will be regulated with a control law for the rate of progression of the virtual target along the path. Having established $\bm{R}_D^T$, we next define $\tilde{\bm{R}}$ as the orientation of the $\mathcal{D}$ frame with respect to $\mathcal{W}$, i.e.,
\begin{equation*}
	\tilde{\bm{R}} \triangleq \bm{R}_{W}^{D} = \bm{R}_{T}^{D}\bm{R}_{W}^{T} = (\bm{R}_{D}^{T})^\top \bm{R}_{W}^{T} \, 
\end{equation*}
We observe that for the vehicle's velocity to be aligned with $\bm{\hat b}_{1D}$, the $(1,1)$ entry of $\tilde{\bm{R}}_{11}$, i.e., $\tilde{\bm{R}}_{11}$, must be equal to 1. From this, we can define the orientation error as
\begin{equation} \label{eq:attitudeerror}
	\Psi(\bm{\tilde{R}}) \triangleq \frac{1}{2}(1 - \tilde{R}_{11}) = \frac{1}{2}\text{tr}[(\bm{\mathds{I}_3-\bm{\mathit{\Pi_R}}^\top \bm{\mathit{\Pi_R}}})(\bm{\mathds{I}_3-\bm{\tilde{R}}})]
\end{equation}
where $\bm{\mathit{\Pi_R}} = \begin{bmatrix} 0 && 1 && 0 \\ 0 && 0 && 1\end{bmatrix}$. As it is shown in \cite{cichella2011geometric,lee2010geometric}, the dynamics of $\Psi(\bm{\tilde{R}})$ are given by
\begin{equation} \label{eq:attitudeerrordynamics}
	{\dot{\Psi}}(\bm{\tilde{R}}) = \bm{e}_{\bm{\tilde{R}}}^\top \left(\begin{bmatrix} \;q\; \\ r \end{bmatrix} - \bm{\mathit{\Pi_R}} \bm{\tilde{R}}^\top\left(\bm{R_{T}^{D}}{\bm{\omega}}_T +\bm{\omega}_{DT}^D\right)\right)
\end{equation}
where $\bm{\omega}_{DT}^D$ is the angular rate of frame $\mathcal{D}$ with respect to frame $\mathcal{T}$ resolved in $\mathcal{D}$, which is determined by
\begin{equation} \label{eq:angratedtd}
    (\bm{\omega}_{DT}^D)^\wedge = (\bm{R}_D^T)^\top \dot{\bm{R}}_D^T  \, 
\end{equation}
and
\begin{equation} \label{eq:eRdef}
\begin{split} 
\bm{e}_{\bm{\tilde{R}}} & \triangleq \frac{1}{2}\bm{\mathit{\Pi_R}}\left(\left(\bm{\mathds{I}_3}-\bm{\mathit{\Pi_R}}^\top\bm{\mathit{\Pi_R}}\right)\bm{\tilde{R}}-\bm{\tilde{R}}^\top\left(\bm{\mathds{I}_3}-\bm{\mathit{\Pi_R}}^\top\bm{\mathit{\Pi_R}}\right)\right)^\vee \\ & =  \frac{1}{2}\left[\tilde{R}_{13}, \, -\tilde{R}_{12}\right]^\top 
\end{split} 
\end{equation}
\begin{asm} \label{asm:autopilot_bound}
The vehicle is equipped with an inner-loop controller that provides tracking capabilities of feasible pitch- and yaw-rate commands, i.e., $q_c(t)$ and $r_c(t)$, respectively. In other words, 
\begin{equation}
\left\Vert
\begin{bmatrix}
q_c(t) - q(t) \\
r_c(t) -r(t)
\end{bmatrix}
\right\Vert
\leq \delta_\omega 
\, , \quad \forall t \geq 0
\end{equation}
if $q_c(t)$ and $r_c(t)$ satisfy
\begin{equation} \label{eq:commandsconstraints}
\left\Vert
\begin{bmatrix}
q_c(t) \\
r_c(t)
\end{bmatrix}
\right\Vert
\leq \omega_{c,\max} 
\, , \quad \forall t \geq 0
\end{equation}
for some $\omega_{c,\max}>0$.
\end{asm} 
With this setup, the path-following problem is stated as follows.
\begin{prob}
\label{prob:path_following}
Derive control laws for $q_c(t)$, $r_c(t)$, and for the rate of progression of the virtual time, $\dot{\gamma}(t)$, such that the path-following error $\bm{e}_{\text{PF}}(t) = [\bm{p_T}^\top (t) , \Psi(\bm{\tilde R (t)})]^\top$ converges to a neighborhood of the origin.  
\end{prob}

\subsection{Path-Following Solution}
\label{sec:pf_solution}
To address the path-following problem we let the dynamics of the virtual time be governed by the  following control law
\begin{equation} \label{eq:gammad}
	\dot{\gamma} = \frac{\left[{v}\bm{\hat{w}_1}+k_\gamma (\bm{p}-\bm{p_d}(\gamma))\right]^\top		\bm{\hat{t}}_{1}(\gamma)}{\left\Vert\bm{p_d'}(\gamma)\right\Vert} \, 
\end{equation}
and let the pitch- and yaw-rate commands be given by
\begin{equation} \label{eq:qcomm}
 	\bm{\omega_c} \triangleq
 	\begin{bmatrix} \; q_c \; \\ \; r_c \; \end{bmatrix}
 	= \bm{\mathit{\Pi_R}} \bm{\tilde R}^\top \left( \bm{R}_{T}^{D}\bm{\omega}_T + \bm{\omega}_{DT}^{D}\right) - 2k_{\tilde{R}} \bm{e_{\tilde{R}}} \, 
\end{equation}
where $k_\gamma > 0$ and $k_{\tilde{R}} >0$ are control gains.

For given angular-rate command constraint $\omega_{c,\max}$ introduced in \eqref{eq:commandsconstraints}, the angular rates ${\bm{\omega}}_T$ and ${\bm{\omega}}_{DT}^D$ defined in \eqref{eq:omegaT} and \eqref{eq:angratedtd} satisfy
$$ \omega_{c, \max}-\omega_{T,\max}\dot{\gamma}-\sup_{t \geq 0} \left\Vert {\bm{\omega}}_{DT}^D \right\Vert > 0 $$
Satisfaction of the above inequality can be ensured by making $\omega_{T,\max}$ and $\sup_{t \geq 0} \left\Vert {\bm{\omega}}_{DT}^D \right\Vert$ arbitrarily small. This can be obtained by tuning the trajectory generation algorithm (see Equation \eqref{eq:boundsTT}) and by increasing the desired parameter $d$ (see Equation \eqref{eq:angratedtd}), respectively. For more details on the satisfaction of the above inequality, see \cite{rober2021three}. Furthermore, let 
\begin{equation} \label{eq:paramc}
\begin{split}
    & c < \min \left\{ \frac{1}{\sqrt{2}} , \frac{1}{2k_{\tilde R}} \left(\omega_{c, \max}-\omega_{T,\max}\dot{\gamma}-\sup_{t \geq 0} \left\Vert {\bm{\omega}}_{DT}^D \right\Vert \right) \right\} \\ & c_1 > 0
\end{split}
\end{equation}
\begin{equation} \label{eq:paramlambda}
    \lambda <  \frac{v_{\min}}{c_1^2 \sqrt{d^2+c^2c_1^2}} , \quad 0< \delta_\lambda < 1 \, 
\end{equation}
The main result is summarized in the following theorem. 
\begin{thm} \label{thm1}
Consider a vehicle moving with speed $v$ satisfying 
\begin{equation} \label{eq:speedasm} 
    0< v_{\min} \leq v \leq v_{\max} 
\end{equation}
Assume that the vehicle is equipped with an inner-loop autopilot satisfying Assumption \ref{asm:autopilot_bound} with 
\begin{equation} \label{eq:deltaomega}
\delta_\omega < 2 \lambda \delta_\lambda c  
\end{equation}
There exist control parameters $d$, $k_\gamma$, and $k_{\tilde{R}}$ such that, for any initial state $\bm{e}_{\text{PF}}(0) \in \Omega_{\text{PF}}$, with
 \begin{equation} \label{eq:domainofattraction}
     \Omega_{\text{PF}} = \left\{ \bm{e}_{PF} : \Psi(\bm{\tilde R}) + \frac{1}{c_1^2} \Vert \bm{p_T} \Vert^2 \leq c^2 \right\} 
 \end{equation}
the rate of progression of the virtual time \eqref{eq:gammad} and the angular-rate commands \eqref{eq:qcomm} ensure that the path-following error $\bm{e}_{\text{PF}}(t)$ is locally uniformly ultimately bounded, and the following bounds hold:
\begin{equation}
    \begin{split}
        & \Psi (\tilde{\bm{R}}(t)) + \frac{1}{c_1^2} \Vert \bm{p_T}(t) \Vert^2 \leq \\ & \qquad  e^{-2 \lambda (1-\delta_\lambda)t}\left( \Psi (\tilde{\bm{R}}(0)) + \frac{1}{c_1^2} \Vert \bm{p_T}(0) \Vert^2 \right)
    \end{split} 
\end{equation}
for all $0 \leq t < T_b$  
\begin{equation} 
    \begin{split} 
\Psi (\tilde{\bm{R}}(t)) + \frac{1}{c_1^2} \Vert \bm{p_T}(t) \Vert^2  \leq \frac{c}{2} \frac{\delta_\omega}{ \lambda \delta_\lambda} , \quad \text{for all} \quad  t \geq T_b 
    \end{split}
\end{equation}
\end{thm}
\textbf{Proof:} The proof of Theorem \ref{thm1} is given in \cite{rober2021three}.

\section{Inner-Loop Problem}
\label{sec:il_problem}
\begin{figure}
    \centering
    \includegraphics[width=0.5\textwidth]{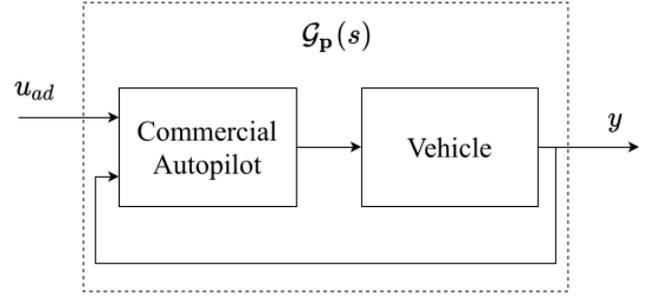}
    \caption{Internals of the system to be controlled: the vehicle packaged with a commercial autopilot}
    \label{fig:simple_block}
\end{figure}

Having addressed the problem of determining angular-rate commands to allow the vehicle to follow a desired path, we next formulate the inner-loop control problem with the goal of finding a system input which allows the vehicle to track the angular-rate commands given by the path-following controller. Namely, we make use of a control strategy similar to the one used in \cite{kaminer2010path} in which an $\mathcal{L}_1$ control algorithm is used to determine the input to a commercial autopilot, shown in Fig. \ref{fig:simple_block}, that is capable of stabilizing the vehicle and providing basic tracking commands.
\begin{asm}
\label{asm:commercial_autopilot}
The vehicle is equipped with an autopilot capable of stabilizing the vehicle and providing tracking capabilities for the vehicle's pitch- and yaw-rates.
\end{asm}
This augmented control architecture improves the autopilot by providing guaranteed performance bounds and behaving more consistently in a variety of situations. These points will be discussed in more detail later. 

As a preliminary to the formulation of the inner loop controller, we must first introduce the $\mathcal{L}_1$ and $\mathcal{L}_{\infty}$ norms as follows:
\begin{equation*}
    \|G(s)\|_{\mathcal{L}_1} = \int_0^\infty|g(t)|dt, \quad \|\bm{\xi}\|_{\mathcal{L}_{\infty}} = \max_{i=1,...n}\left(\sup_{\tau \geq 0}|\xi_i(\tau)|\right)
\end{equation*}
Also note that we make use of the right psudo-inverse of a full row-rank matrix $\bm{T} \in \mathds{R}^{m \times n}$, which is denoted by $\bm{T}^\dagger$ and is calculated as $\bm{T}^\dagger=\bm{T}^\top\left(\bm{T}\bm{T}^\top\right)^{-1}$, thus giving $\bm{T}\bm{T}^\dagger=\mathds{I}_m$

\subsection{Inner-Loop Problem Formulation}
To begin mathematically formulating the inner-loop control structure, we first introduce the autopilot-vehicle system as multi-input multi-output system:
\begin{equation}
    \bm{\mathcal{G}_p}(s)
    \begin{cases}
    \label{eq:autopilot}
        \dot{\bm{x}}(t) = \bm{A_{p}} \bm{x}(t) + \bm{B_{p}} (\bm{u_{ad}}(t)+\bm{f}(t,\bm{x}(t))), \quad \bm{x}(0) = \bm{x}_{0}\\
        \bm{y}(t) = \bm{C_{p} x}(t) \\
    \end{cases}
\end{equation}
where $\bm{u_{ad}}(t) \in \mathds{R}^2$ is the input consisting of pitch- and yaw-rate reference signals, i.e., $\bm{u_{ad}}\triangleq [q_{ad},\;r_{ad}]^{\top}$, $\bm{y}(t) \in \mathds{R}^2$ is the output containing the vehicle's actual pitch- and yaw-rates, i.e., $\bm{y}(t) \triangleq [q(t),\;r(t)]^{\top}$, and $\bm{f}(t,\bm{x}(t))\in \mathds{R}^2$ represents the time-varying uncertainties and nonlinearities associated with the system. $\left\{\bm{A_{p}} \in \mathds{R}^{n_{p} \times n_{p}},\;\bm{B_{p}} \in \mathds{R}^{n_{p} \times 2},\;\bm{C_{p}} \in \mathds{R}^{2 \times n_{p}}\right\}$ is a known controllable-observable triple with the unknown initial condition $\bm{x}_{0}\in \mathds{R}^{n_{p}}$ which is assumed to be inside the arbitrarily large set $\|\bm{x}_0\|_{\infty} < \rho_0 < \infty$ for some known $\rho_0>0$.

\begin{rem}
Note that the triple $\left\{\bm{A_{p}},\;\bm{B_{p}},\;\bm{C_{p}}\right\}$ is assumed to be known only in the context of the theoretical formulation of performance bounds. It is not used in the formulation or implementation of the controller.
\end{rem}

Next we introduce the desired systems $\bm{M}(s)$ with $$\left\{\bm{A_{m}} \in \mathds{R}^{n_{m} \times n_{m}},\;\bm{B_{m}} \in \mathds{R}^{n_{m} \times 2},\;\bm{C_{m}} \in \mathds{R}^{2 \times n_{m}}\right\}$$ i.e.,
\begin{equation*}
    \bm{M}(s) \triangleq \bm{C_{m}}\left(s\mathds{I}-\bm{A_{m}}\right)^{-1}\bm{B_{m}}
\end{equation*}
$\bm{M}(s)$ is a design parameter of the $\mathcal{L}_1$ controller and specifies $\bm{y_m} \triangleq [q_m,\;r_m]^{
\top}$ where $q_m$ and $r_m$ are the desired pitch- and yaw-rate behaviors in response to the commands $q_c$ and $r_c$. The triple $\left\{\bm{A_{m}},\;\bm{B_{m}},\;\bm{C_{m}}\right\}$ must be selected such that $\bm{A_m}$ is Hurwitz, $\bm{C_m}\bm{B_m}$ is nonsingular, and $\bm{M}(s)$ does not have a non-minimum-phase transmission zero. The desired dynamics are given by the Laplace transform
\begin{equation}
\label{eq:desired}
    \bm{y_m}(s) = \bm{M}(s) \bm{K_{g}} \bm{\omega_c}(s), \quad \bm{K_{g}} = -(\bm{C_{m}} \bm{A_{m}}^{-1} \bm{B_{m}})^{-1}
\end{equation}
where $\bm{\omega_c}(s)$ is the Laplace transform of the angular-rate command given in Eq. \eqref{eq:qcomm}.
Thus, the problem becomes a matter of designing a control law for $\bm{u_{ad}}$ such that the output $\bm{y}$ of system \eqref{eq:autopilot} tracks the desired output $\bm{y_m}$ given in Eq. \eqref{eq:desired}.

To aid in the solution of this problem, we first introduce the following series of assumptions.
\begin{asm}
\label{asm:bounded_input}
The reference signal $\bm{\omega_c}(t)$ given by the path-following algorithm is bounded such that
\begin{equation}
\label{eq:discrete_command}
    \| \bm{\omega_c}(t) \|_{\infty} \leq M_{\omega}
\end{equation}
for some known constant $M_{\omega}>0$.
\end{asm}
Notice that Assumption \ref{asm:bounded_input} differs from the bound given in Eq. \eqref{eq:commandsconstraints} of Assumption \ref{asm:autopilot_bound} which uses the Euclidean norm.
\begin{asm}
\label{asm:uncertainty}
For any $\delta>0$ there exists $F_{\delta}$ and $L_0$ such that
\begin{align*}
    \| \bm{f}(t,\bm{x}_2) - \bm{f}(t,\bm{x}_1) \|_{\infty} &\leq F_{\delta} \|\bm{x}_2 - \bm{x}_1 \|_{\infty}, \quad \\
    \|\bm{f}(t,\bm{0}) \|_{\infty} &\leq L_0
\end{align*}
hold uniformly for all $\|\bm{x}_i\|_{\infty}< \delta, i \in \{1,2\}, t \geq 0$.
\end{asm}
With these assumptions in place, we formally state the problem of developing the inner-loop autopilot augmentation algorithm below.
\begin{prob}
\label{prob:inner_loop}
Find a control law for $\bm{u_{ad}}(t)$ such that the pitch- and roll-rate of the vehicle's flow frame, packaged into $\bm{y}(t)$, exhibits the desired behavior given by Eq. \eqref{eq:desired}.
\end{prob}

\subsection{Inner-Loop Structure}
\label{sec:inner_loop_solution}
\begin{figure}
\centering
\includegraphics[width=.47\textwidth]{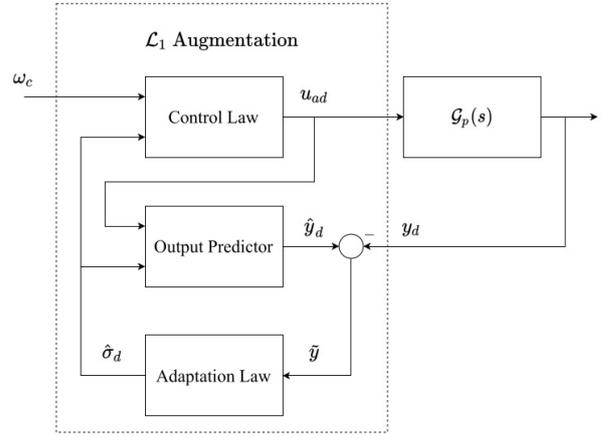}
\caption{Structure of $\mathcal{L}_1$ augmentation loop, adapted from \cite{kaminer2010path}}
\label{fig:structure}
\end{figure}

To determine the input $\bm{u_{ad}}(t)$ which controls the system given in Eq. \eqref{eq:autopilot} to behave like the desired system given by Eq. \eqref{eq:desired}, we use a set of discrete functions comprising of $(i)$ an adaptation law that estimates the value of $\bm{f}(t,\bm{x})$, $(ii)$ an output predictor that uses the estimate from $(i)$ and the previous value of $\bm{u_{ad}}$ to predict the value of $\bm{y}$, and $(iii)$ a control law that uses the estimate from $(i)$ with the commands from the path-following controller to determine the value of $\bm{u_{ad}}$ that solves Problem \ref{prob:inner_loop}. A block diagram of this structure is shown in Fig. \ref{fig:structure}.

\subsubsection{Adaptation Law}
Let $\bm{P}=\bm{P}^{\top}>0$ be the solution to the Lyapunov equation 
\begin{equation*}
    \bm{A_{m}}^{\top}\bm{P}+\bm{P}\bm{A_{m}} = -\bm{Q},\quad \bm{Q} = \bm{Q}^{\top}>0
\end{equation*}
The properties of $\bm{P}$ imply that there exists a nonsingular matrix $\sqrt{\bm{P}}$ such that 
\begin{equation*}
    \bm{P} = \sqrt{\bm{P}}^{\top}\,\sqrt{\bm{P}}
\end{equation*}
Then, given the vector $\bm{C_{m}} \sqrt{\bm{P}}^{-1}$, let $\bm{D}$ be the $\left((n_{m}-1) \times n_{m}\right)$ nullspace of $\bm{C_{m}} \sqrt{\bm{P}}^{-1}$, i.e.,
\begin{equation*}
    \bm{D}\left(\bm{C_{m}} \sqrt{\bm{P}}^{-1}\right)^{\top} = 0
\end{equation*}
Define $\bm{\Lambda}$ as 
\begin{equation*}
    \bm{\Lambda} \triangleq 
    \begin{bmatrix}
    \bm{C_{m}} \\ \bm{D} \sqrt{\bm{P}}
    \end{bmatrix}
    \in \mathds{R}^{n_{m} \times n_{m}}
\end{equation*}
The purpose of the adaptation law is to provide a discrete-time estimate $\bm{\hat{\sigma}_d}(t)$ of the desired-system nonlinearity term $\bm{f}(t,\bm{x}(t))$. To formulate this, we write 
\begin{equation*}
    \bm{\hat{\sigma}_d}(t) = \bm{\hat{\sigma}_d}[i],\quad t \in [iT_s,(i+1)T_s),\quad i \in \mathds{Z}_{\geq 0}
\end{equation*}
The discrete-time nonlinearity estimate at a given time step is determined by
\begin{equation}
\label{eq:adaptation_law}
    \bm{\hat{\sigma}_d}[i] = - \bm{\Phi}^{-1}(T_s)e^{\bm{\Lambda} \bm{A_{m}} \bm{\Lambda}^{-1}T_s}\bm{1}_{n_m2}(\bm{\hat{y}_d}[i]-\bm{y_d}[i])
\end{equation}
where $\bm{1}_{n_m2} \triangleq \left[\mathds{I}_2, \; 0_{2 \times (n_m-2)} \right]^{\top} \in \mathds{R}^{n_m \times 2}$, $\bm{y_d}[i]$ is the sampled output i.e., $\bm{y_d}[i]=\bm{y}(iT_s)$, $\bm{\hat{y}_d}[i]$ is the discrete output estimate given by the output predictor, and $\bm{\Phi}(T_s)$ is an $\left(n_{m} \times n_{m}\right)$ matrix given by
\begin{equation*}
    \bm{\Phi}(T_s) = \int_0^{T_s} e^{\bm{\Lambda} \bm{A_{m}} \bm{\Lambda}^{-1}\left(T_s-\tau\right)}\bm{\Lambda} d\tau
\end{equation*}
Note that $\bm{\sigma_d}[i]$ is determined with a $T_s$-dependent constant multiplied by the error in the output measurement. This means that $T_s$ is the only factor determining the speed of the adaptation law and that it can be made arbitrarily fast by sufficiently reducing $T_s$.

\subsubsection{Output Predictor}
We employ the discrete output predictor given by
\begin{equation}
\label{eq:output_predictor}
\begin{split} 
    \bm{\hat{x}_d}[i+1] & = e^{\bm{A_{m}}T_s}\bm{\hat{x}_d}[i] \\ & \qquad +\bm{A_{m}}^{-1}\left(e^{\bm{A_{m}} T_s}-\mathds{I}_{n_m}\right)\left(\bm{B_{m}} \bm{u_d}[i]+\bm{\hat{\sigma}_d}[i]\right), \\  \bm{\hat{y}_d}[i] & = \bm{C_{m}}\bm{\hat{x}_d}[i],\quad \bm{x_d}[0] = \bm{C_{m}}^{\dagger}\bm{y}_0
\end{split}
\end{equation}
where $\bm{\hat{x}_d}[i] \in \mathds{R}^{n_{m}}$ is the state, $\bm{u_d}[i] \in \mathds{R}$ is the input determined by the control law, and $\bm{y}_0 = \bm{y}(0)$ is the known initial output value. Note that the output predictor replicates the desired closed-loop dynamics, but substitutes the unknown function $\bm{f}(t,\bm{x}(t))$ with $\bm{\hat{\sigma}_d}[i]$.

\subsubsection{Control Law}
The system input $\bm{u_{ad}}$ is first discretized as
\begin{equation*}
    \bm{u_{ad}}(t) = \bm{u_d}[i],\quad t \in [iT_s,(i+1)T_s),\quad i \in \mathds{Z}_{\geq 0}
\end{equation*}
where $\bm{u_d}[i]$ is determined by
\begin{equation}
\label{eq:control_law}
\begin{split}
    \bm{x_u}[i+1] & = e^{\bm{A_o}T_s}\bm{x_u}[i] \\ & \qquad +\bm{A_o}^{-1}\left(e^{\bm{A_o} T_s}-\mathds{I}_{n_o}\right)\left(\bm{B_o}e^{-\bm{A_{m}}T_s}\bm{\hat{\sigma}_d}[i]\right), \\
    \bm{u_d}[i] & = \bm{K_{g}}\bm{\omega_c}[i]-\bm{C_o}\bm{x_u}[i],\quad \bm{x_u}[0] = 0
\end{split}
\end{equation}
where $\bm{K_{g}}$ and $\bm{\omega_c}[i]$ were introduced in Eqs. \eqref{eq:desired} and \eqref{eq:discrete_command}, and $(\bm{A_o} \in \mathds{R}^{n_o \times n_o},\;\bm{B_o}  \in \mathds{R}^{n_o \times 2},\;\bm{C_o}  \in \mathds{R}^{2 \times n_o})$ is the minimal state-space realization of the transfer function
\begin{equation}
\label{eq:filter}
    \bm{O}(s) = \bm{C}(s)\bm{M}^{-1}(s)\bm{C_{m}}\left(s\mathds{I}_{n_o}-\bm{A_{m}}\right)^{-1}
\end{equation}
and $\bm{C}(s)$ is a strictly proper stable transfer function such that $\bm{C}(0) = \mathds{I}_2$.

In summary, the inner-loop control algorithm is comprised of Eqs. (\ref{eq:adaptation_law}-\ref{eq:filter}). To analyze the stability of the given controller, it is necessary to introduce a series of terms that will help to define stability conditions and set requirements for the controller's design:
\begin{equation}
    \begin{split}
        \bm{P}(s) & \triangleq \bm{C_p}\left(s \mathds{I}_{n_p} - \bm{A_p} + \bm{B_p} \bm{K}\right)^{-1} \bm{B_p} \\
        \bm{H_0}(s) & \triangleq \left(s \mathds{I}_{n_p} - \bm{A_p} + \bm{B_p} \bm{K}\right)^{-1}\bm{B_{p}} \\
        \bm{H_1}(s) & \triangleq (\mathds{I}_2+\left(\bm{M}^{-1}(s)\bm{P}(s)-\mathds{I}_2)\bm{C}(s)\right)^{-1} \\
        \bm{H_2}(s) & \triangleq \bm{H_0}(s)-\bm{H_0}(s)\bm{C}(s)\bm{H_1}(s)\left(\bm{M}^{-1}(s)\bm{P}(s)-\mathds{I}_2\right) \\
        \bm{H_3}(s) & \triangleq \bm{H_1}(s)\bm{M}^{-1}(s)\bm{P}(s) \\
        \bm{H_4}(s) & \triangleq \bm{H_1}(s)\left(\bm{M}^{-1}\bm{P}(s)-\mathds{I}_2\right) \\
        \bm{H_5}(s) & \triangleq \bm{H_0}(s)\bm{C}(s)\bm{H_1}(s)\bm{M}^{-1}(s) \\
        \bm{G}(s) & \triangleq \bm{H_0}(s) - \bm{H_5}(s)\bm{P}(s) \\
        \bm{H_{\mathrm{in}}}(s) & \triangleq \bm{C_p}\left(s\mathds{I}_{n_p}-\bm{A_p}+\bm{B_p}\bm{K}\right)^{-1} \\ & \qquad -\bm{C_m}\left(s\mathds{I}_{n_m}-\bm{A_m}\right)^{-1}\bm{C_m}^{\dagger}\bm{C_p}
    \end{split}
\end{equation}
where $\bm{K} \in \mathds{R}^{2 \times n_p}$ is selected such that $\bm{A_p} - \bm{B_p} \bm{K}$ is Hurwitz. With these definitions in place, we reiterate that $\bm{C}(s)$ must be strictly proper with $\bm{C}(0) = \mathds{I}_2$ and add that $\bm{C}(s)$ must be chosen such that
\begin{equation} \label{eq:cond1}
    \bm{H_1}(s)\;\mathrm{is\;stable}
\end{equation}
\begin{equation} \label{eq:cond2}
    \bm{C}(s)\bm{M}^{-1}(s)\; \mathrm{is\;proper}
\end{equation}
and for some $\rho_0$ there exists $\rho_r>\rho_0$ satisfying
\begin{equation} \label{eq:cond3}
    \|\bm{G}(s)\|_{\mathcal{L}_1} < \frac{\rho_r-\rho_1-\rho_2}{L_{\rho_r}\rho_r+L_0}
\end{equation}
where
\begin{equation} 
    \rho_1 \triangleq \left\lVert s\left(s\mathds{I}_{n_p}-\bm{A_p}+\bm{B_p}\bm{K}\right)^{-1}-s\bm{H_5}(s)\bm{H_{\mathrm{in}}}(s)\right\rVert_{\mathcal{L}_1}\rho_0 
\end{equation}
\begin{equation} 
    \rho_2 \triangleq  \left\lVert \bm{H_2}(s)\bm{K_g}\right\rVert_{\mathcal{L}_1}M_{\omega}
\end{equation}
and 
\begin{equation}
    L_{\rho_r} \triangleq \frac{\bar{\gamma}_1+\rho_r}{\rho_r}\left(F_{(\delta+\bar{\gamma}_1)}+\|\bm{K}\|_{\infty}\right)
\end{equation}
where $\bar{\gamma}_1$ is a small constant and $F_{\delta}$ was introduced in Assumption \ref{asm:uncertainty}.

Next we introduce the reference system
\begin{equation} \label{eq:reference_system}
    \begin{split}
        \bm{\dot{x}_{\mathrm{ref}}}(t) = & \: \bm{A_p}\bm{x_{\mathrm{ref}}}(t)+\bm{B_p}\left(\bm{u_{\mathrm{ref}}}(t)+\bm{f}(t,\bm{x_{\mathrm{ref}}}(t))\right) \\
        \bm{u_{\mathrm{ref}}}(s) = & \: \bm{K_g}\bm{\omega_c}(s) - \bm{C}(s)\bm{\sigma_{\mathrm{ref}}}(s) \\
        \bm{y_{\mathrm{ref}}}(t) = & \: \bm{C_p}\bm{x_{\mathrm{ref}}}(t),\quad \bm{x_{\mathrm{ref}}}(0) = \bm{x}_0
    \end{split}
\end{equation}
where
\begin{equation}
\begin{split} 
    \bm{\sigma_{\mathrm{ref}}}(s) & = \Big( \left[\left(\bm{P}(s)-\bm{M}(s)\right)\bm{C}(s)+\bm{M}(s)\right]^{-1} \\ & \qquad \left(\bm{P}(s)-\bm{M}(s)\right)\bm{K_g}\bm{\omega_c}(s)\Big) \\ & \qquad \quad +\Big(\left[\left(\bm{P}(s)-\bm{M}(s)\right)\bm{C}(s)+\bm{M}(s)\right]^{-1} \\ & \qquad \left(\bm{P}(s)\bm{\nu_{\mathrm{ref}}}(s)+\bm{H_{\mathrm{in}}}(s)\bm{x}_0\right) \Big)
\end{split}
\end{equation}
and $\bm{\nu_{\mathrm{ref}}}(s)$ is the Laplace transform of
\begin{equation}
    \bm{\nu_{\mathrm{ref}}}(t) = \bm{K}\bm{x_{\mathrm{ref}}}(t)+\bm{f}(t,\bm{x_{\mathrm{ref}}}(t))
\end{equation}
Notice that the output from Eq. \eqref{eq:reference_system} can be reformulated to show that
\begin{equation} \label{eq:reference_output}
\begin{split}
    \bm{y_{\mathrm{ref}}}(s) & = \bm{M}(s)\bm{K_g}\bm{\omega_c}(s)+\bm{M}(s)\left(\mathds{I}_2-\bm{C}(s)\right)\bm{\sigma_{\mathrm{ref}}}(s) \\ & \qquad +\bm{C_m}\left(s\mathds{I}_2-\bm{A_m}\right)^{-1}\bm{C_m}^{\dagger}\bm{y}_0
\end{split}
\end{equation}
which highlights the fact that the uncertainty $\bm{\sigma_{\mathrm{ref}}}(t)$ can only be reduced within the bandwidth of $\bm{C}(s)$. Additionally, because $\bm{C}(0) = \mathds{I}_2$, the Final Value Theorem applied to Eq. \eqref{eq:reference_output} implies that the reference output follows the desired output given by Eq. \eqref{eq:desired}.
The reference system given by Eq. \eqref{eq:reference_system} serves as the ideal form of the system given by Eq. \eqref{eq:autopilot} controlled by Eqs. (\ref{eq:adaptation_law}-\ref{eq:filter}) which is obtained as $T_s$ approaches 0 and serves to provide a measurement of the performance of the real system. Thus, the next lemma establishes the stability and performance bounds of the reference system, which will later be used to define the real system's performance.

\begin{lem} 
\label{lem1}
Consider the closed-loop reference system given in \eqref{eq:reference_system} subject to (\ref{eq:cond1}-\ref{eq:cond2}). If $\|\bm{x}_0\|_{\infty} \leq \rho_0$, then
\begin{equation}
    \|\bm{x_{\mathrm{ref}}}\|_{\mathcal{L}_1} < \rho_r
\end{equation}
\begin{equation}
    \|\bm{u_{\mathrm{ref}}}\|_{\mathcal{L}_1} < \rho_{ur}
\end{equation}
where $\rho_{ur}$ is given by
\begin{equation}
\begin{split} 
    \rho_{ur} & \triangleq \left\lVert \bm{C}(s)\bm{H_3}(s) \right\rVert_{\mathcal{L}_1} \left(L_{\rho_r}\rho_r+L_0\right) \\ & \qquad +\left\lVert s\bm{C}(s)\bm{H_1}(s)\bm{M}^{-1}(s)\bm{H_{\mathrm{in}}}(s) \right\rVert_{\mathcal{L}_1}\rho_0 \\ & \qquad + \left\lVert \left(\mathds{I}_2-\bm{C}(s)\bm{H_4}(s)\right)\bm{K_g} \right\rVert_{\mathcal{L}_1}M_{\omega}
\end{split} 
\end{equation}
\end{lem}
The proof for Lemma \ref{lem1} can be found in \cite{jafarnejadsani2018robust}

Finally, in Theorem \ref{thm2}, we define the performance bounds of the control structure given by Eqs. (\ref{eq:adaptation_law}-\ref{eq:filter}) applied to the closed-loop autopilot system given by Eq. \eqref{eq:autopilot} relative to the reference system in Eq. \eqref{eq:reference_system}.

\begin{thm} \label{thm2}
Consider the system given in Eq. \eqref{eq:autopilot} with the control laws given by Eqs. (\ref{eq:adaptation_law}-\ref{eq:filter}) subject to conditions \eqref{eq:cond1}, \eqref{eq:cond2}, and \eqref{eq:cond3}. If $\|\bm{x}_0\|_{\infty} < \rho_0$, then
\begin{equation*}
    \|\bm{x_{\mathrm{ref}}}(t)-\bm{x}(t)\|_{\mathcal{L}_{\infty}} \leq \gamma_x
\end{equation*}
and
\begin{equation*}
    \|\bm{u_{\mathrm{ref}}}(t)-\bm{u_{ad}}(t)\|_{\mathcal{L}_{\infty}} \leq \gamma_{u_{ad}}
\end{equation*}
with
\begin{equation*}
    \lim_{T_s \to 0} \gamma_x = \lim_{T_s \to 0} \gamma_{u_{ad}} = 0
\end{equation*}
\end{thm}
The proof for Theorem \ref{thm2} and more detail on the definitions of $\gamma_x$ and $\gamma_{u_{ad}}$ can be found in \cite{jafarnejadsani2018robust}.

Theorem \ref{thm2} implies that the behavior of the system given by Eq. \eqref{eq:autopilot} can be made arbitrarily close to the behavior of the reference system given by Eq. \eqref{eq:reference_system} by reducing the sampling rate $T_s$. Moreover, as is shown by Eq. \eqref{eq:reference_output}, $\bm{y_{\mathrm{ref}}}$ tracks the desired output $\bm{y_m}$. Thus, from Theorem \ref{thm2}, it can be concluded that the output $\bm{y}$ tracks the desired output $\bm{y_m}$ uniformly in transient and steady state with performance bounds that can be decreased through the selection of $T_s$ and $\bm{C}(s)$.

The sampling time $T_s$ acts as an adaptation gain, and while it's not necessary that it matches the CPU clock cycle time, the lowest possible $T_s$ is desirable as it tightens the error bounds between the reference system and the actual system as specified by Theorem \ref{thm2}. The control parameter $\bm{C}(s)$ acts as a low-pass filter which attenuates the high-frequency components of the adaptation law. It thus acts as a design variable which can be used to tune the balance between the sensitivity of the adaptation law to disturbances and the robustness of the controller. Further discussion on the selection of $\bm{C}(s)$ can be found in \cite{hovakimyan2010} and \cite{jafarnejadsani2018robust}

\begin{rem}
As previously stated, use of the Final Value Theorem on Eq. \eqref{eq:reference_output} shows that $\bm{y_m} \to \bm{y_{\mathrm{ref}}}$ as $T_s \to 0$. Combining this with the result from Theorem \ref{thm2}, it can be shown that $\|\bm{y_m}-\bm{y}\|_{\mathcal{L}_\infty}$ is bounded. Moreover, because $\bm{M}(s)$ is bounded-input, bounded-output stable, it follows that $\|\bm{\omega_c}-\bm{y}\|_{\mathcal{L}_\infty}$ is bounded. This result can be applied to Assumption \ref{asm:autopilot_bound} to find path-following error bounds as they relate to the inner-loop control parameters.
\end{rem}

\section{Simulation Results} \label{sec:results}
In this section we provide numerical results demonstrating the previously described control scheme on an underwater vehicle modeled in Simulink. The Simulink model used is a reduced-order hydrodynamic model which uses coefficients from a series of CFD experiments to model the inertial forces and hydrodynamic loads experienced by the Joubert BB2 underwater vehicle introduced in \cite{carrica2019}. The hydrodynamic forces captured by this model include surface effects, i.e., the hydrodynamic forces caused by motion of a body near the free surface or the sea floor, which can cause a significant change in the hydrodynamic load throughout a maneuver which experiences a change in depth. The development of this Simulink model is documented in \cite{kim2021development} along with specific details on the model's properties. We will present three sets of maneuvers: (1) a depth-change, (2) a combined depth- and $y_W$-position-change, and (3) a canyon-traversal.

Each of the maneuvers employs Bernstein polynomials (BPs) as the basis for the desired path. Thus, $\bm{p_d}$ takes the form
\begin{equation*}
    \bm{p_d}(\gamma(t))  
    = \sum_{j=0}^N \bm{\bar{p}}_{j} b_{j,N}(\gamma(t))
\end{equation*}
where $\bm{\bar{p}}_{0},\ldots,\bm{\bar{p}}_{N} \in \mathds{R}^3$ are the Bernstein coefficients, and $b_{j,n}(\gamma(t)) = \binom{N}{j} \frac{\gamma(t)^j(T_f-\gamma(t))^{N-j}}{T_f^N} $ is the Bernstein polynomial basis with $\binom{N}{j} = \frac{N ! }{j ! (N-j)!}$. Selecting BPs as the basis for $\bm{p_d}$ provides two advantages: ($i$) BPs are continuous, ensuring the satisfaction of the path-following condition described by Eq. \eqref{eq:commandsconstraints}, and ($ii$) optimization techniques can be used to generate approximately optimal trajectories as is demonstrated in \cite{cichella2019optimal}. For the purpose of this paper, they allow us to select Bernstein coefficients to create arbitrary paths.

\begin{figure}
    \centering
    \includegraphics[width=1\linewidth]{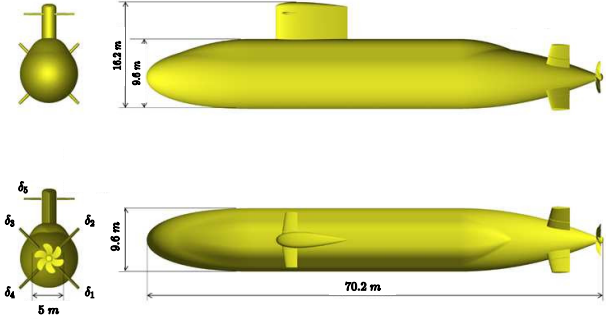}
    \caption{Joubert BB2 model highlighting x-shaped stern plane configuration, adapted from \cite{carrica2019}}
    \label{fig:joubert}
\end{figure}

In order to satisfy Assumption \ref{asm:commercial_autopilot}, we must design an autopilot controller capable of stabilizing the vehicle and tracking some reference command. The x-shaped configuration of the vehicle's stern planes, shown in Fig. \ref{fig:joubert}, requires that each control surface has a unique command to allow the submarine to maneuver in 3D space. Determination of each control surface command is given by
\begin{equation*}
    \delta_i =
    \begin{cases}
     \bar{\delta}_i \, \qquad \delta_{\min} \leq \bar{\delta}_i \leq \delta_{\max} \\ 
     \delta_{\min} \, \qquad \bar{\delta}_i \leq \delta_{\min} \\
     \delta_{\max} \, \qquad \bar{\delta}_i \geq \delta_{\max} \\
    \end{cases}
\end{equation*}
where $\delta_{\max}=-\delta_{\min}=30$ degrees, and $\bar{\delta}_i,\;i \in \{1,2,3,4,5\}$ is given by
\begin{align*}
    && \bar{\delta}_1 = -\delta_v+\delta_h
    && \bar{\delta}_3 = \delta_v-\delta_h
    && \bar{\delta}_5 = \delta_v \\
    && \bar{\delta}_2 = -\delta_v-\delta_h
    && \bar{\delta}_4 = \delta_v+\delta_h
\end{align*}
with $\delta_v$ and $\delta_h$ given by the PI control laws
\begin{align*}
    \delta_v & = K_{P_v}\left(\bar{u}_q-q\right)+K_{I_v}\int_0^t\left(\bar{u}_q-q\right)dt \\
    \delta_h & = K_{P_h}\left(\bar{u}_r-r\right)+K_{I_h}\int_0^t\left(\bar{u}_r-r\right)dt
\end{align*}
with control gains $K_{P_v}=K_{P_h}=3000$ and $K_{I_v}=K_{I_h}=50$ and $[\bar{u}_q,\;\bar{u}_r]^\top \triangleq \bm{u_{ad}}$ for cases labeled "Adaptation On" and $[\bar{u}_q,\;\bar{u}_r]^\top \triangleq \bm{\omega_c}$ for cases labeled "Adaptation Off". 
The control parameters used to obtain the results in this section were selected so that the vehicle would quickly converge to the desired path without being overly aggressive. The path-following gains are given by
\begin{align*}
k_\gamma = 1\;[1/\mathrm{s}] && k_{\tilde{R}} = 0.1\;[1/\mathrm{s}] && d = 250\;[\mathrm{m}]
\end{align*}
The adaptive parameters, i.e., the desired system $\bm{M}(s)$ and the low-pass filter $\bm{C}(s)$, were designed to be
\begin{equation} \label{eq:il_pars}
    \bm{M}(s) =
    \begin{bmatrix}
        \frac{0.1}{s+0.1} & 0 \\
        0 & \frac{0.1}{s+0.1}
    \end{bmatrix}
\end{equation}
\begin{equation}
    \bm{C}(s) = 
    \begin{bmatrix}
        \frac{0.1}{(s+1)^2(s+0.1)} & 0 \\
        0 & \frac{0.01^3}{(s+.01)^3}
    \end{bmatrix}
\end{equation}
These must be designed while considering the dynamics of the system and the nature of the disturbances it may experience. Thus, using insight gained from \cite{kim2021development} about the speed of the pitch and yaw dynamics, $\bm{M}(s)$ was designed with moderately fast dynamics. $\bm{C}(s)$ was designed to balance robustness and adaptation such that the pitch-channel is able to more quickly adapt to disturbances such as waves and surface interactions. We set $T_s = 0.05$ except where noted. 

\begin{figure}
    \centering
    \includegraphics[width=0.95\linewidth,trim={0cm 0 0cm 0},clip]{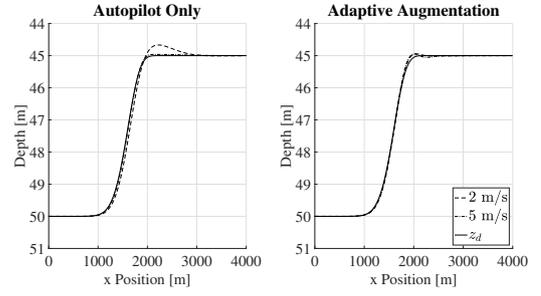}
    \caption{Depth change maneuver executed with adaptation off and on at 2 m/s and 5 m/s}
    \label{fig:depth_change}
\end{figure}

First, we present a simple depth-change maneuver to set a baseline for the controller's performance and compare the adaptive and non-adaptive autopilot performances at both 2 and 5 m/s. The desired path for the maneuver is defined by a BP with control points stepping from 50 m to 45 m in depth, thus providing a smooth step function. The results for this maneuver are shown in Fig. \ref{fig:depth_change}.
The maximum overshoot values without adaptation were $0.35$m and $0.02$m for 2 and 5 m/s respectively, while the corresponding values with adaptation were approximately $0.02$m for both maneuvers. Notice that with only the autopilot, the vehicle still performs well at 5 m/s where it has more control authority. However, in the 2 m/s case, the control surfaces' ability to pitch the vehicle is diminished, resulting in a larger maximum overshoot. In contrast, the adaptive controller is able to estimate the vehicle's dynamics at both 2 and 5 m/s, and controls it in a way that is consistent for both speeds. 

\begin{figure}
    \centering
    \includegraphics[width=1\linewidth]{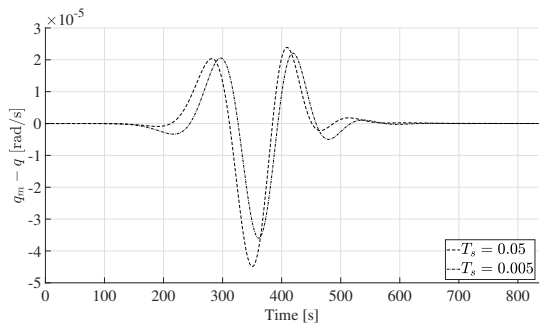}
    \caption{Adaptive tracking error for depth change maneuvers with $T_s=0.05$ s and $T_s=0.005$ s}
    \label{fig:q_comp}
\end{figure}

As stated in Theorem \ref{thm2}, the vehicle's behavior approaches the desired behavior as $T_s$ approaches 0. In Fig. \ref{fig:q_comp}, we compare the tracking error $q_m-q$ for a \emph{slow} $T_s=0.05$ case (already shown in Fig. \ref{fig:depth_change}), and \emph{fast} $T_s=0.005$ case.
Notice that as $T_s$ is reduced, the maximum error between $q_m$ and $q$ is reduced. While some points on the error curve for the slow case are smaller than those of the fast curve, the $\mathcal{L}_\infty$-norm condition in Theorem \ref{thm2} only requires that the maximum error is reduced. Fig. \ref{fig:q_comp} confirms this with the fast case showing a 19.9\% reduction in the maximum error compared to the slow case.

\begin{figure}
\centering
    \begin{subfigure}{0.5\textwidth}
        \includegraphics[width=1\linewidth]{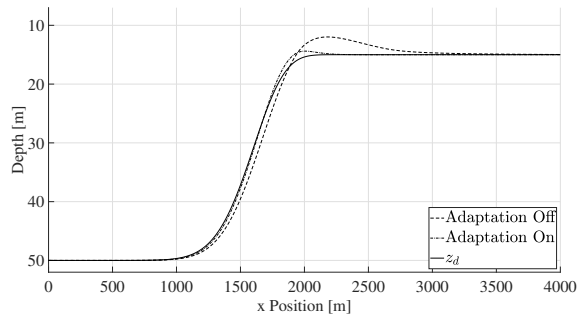}
        \caption{Vertical plane results for lane-change maneuver at 2 m/s}
        \label{fig:lane_change_z}
    \end{subfigure}
    \begin{subfigure}{0.5\textwidth}
        \includegraphics[width=1\linewidth]{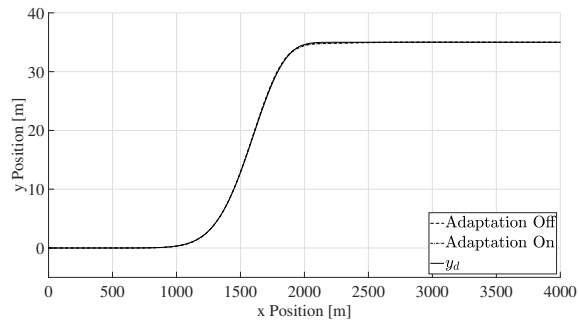}
        \caption{Horizontal plane results for lane-change maneuver at 2 m/s}
        \label{fig:lane_change_y}
    \end{subfigure}
    \caption{Lane-change maneuver}
    \label{fig:lane_change_maneuver}
\end{figure}

Next we show a more challenging \emph{lane-change} maneuver in which the vehicle is commanded to perform a depth change from 50 to 15 m in depth and simultaneously change it's $y_W$ position from 0 to 35 m. This maneuver brings the vehicle closer to the free surface where surface effects introduce an additional suction force, thus adding an environmental disturbance the controller must contend with. To increase the difficulty, this maneuver was executed at a commanded forward speed of 2 m/s. Fig. \ref{fig:lane_change_maneuver} shows the results of this maneuver.
The results in Fig. \ref{fig:lane_change_z} show that, in addition to the tracking error during the transient period, the suction forces from the surface effects cause a greater maximum overshoot for the non-adaptive case (3.02 m for the non-adaptive case vs 0.61 m for the adaptive case). This is notable because the vehicle's sail breaks the surface when its CG is at a depth of 11 m. Breaking the surface is detrimental for stealth operations, thus demonstrating what could be considered a safety-critical maneuver where it is important to have a controller with guaranteed performance bounds. Fig. \ref{fig:lane_change_y} shows that a similar effect does not appear in the $y_{\mathrm{w}}$-position change because the suction forces only affect the vertical plane. Note that the lack of righting moment and comparatively simple horizontal dynamics result in very similar results between the adaptive and non-adaptive cases, demonstrating the shifted balance toward robustness given by the more restrictive filter for the $r$-channel in Eq. \eqref{eq:il_pars}.

\begin{figure}
    \centering
    \includegraphics[width=1\linewidth]{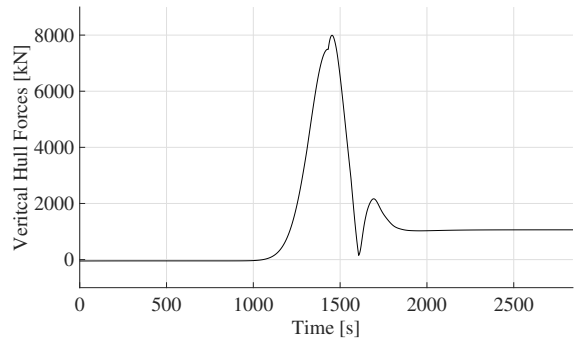}
    \caption{Vertical hydrodynamic force on the hull of the vehicle}
    \label{fig:hull_force}
\end{figure}

\begin{figure}
    \centering
    \includegraphics[width=1\linewidth]{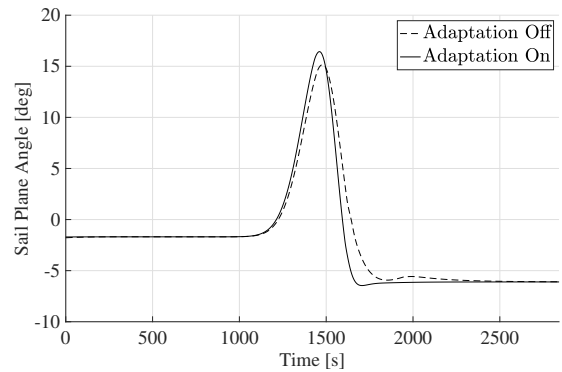}
    \caption{Comparison of the sail plane angle with adaptation on and adaptation off}
    \label{fig:sail_planes}
\end{figure}

To help quantify the magnitude of the disturbance caused by the suction force due to hydrodynamic interaction between the vehicle and the free surface, Fig. \ref{fig:hull_force} shows the vertical hydrodynamic force on the hull of the vehicle during the "Adaptation On" simulation from Fig. \ref{fig:lane_change_maneuver}. Notice that when the vehicle is at a depth of 50 m ($t=0$ to $t=1000$), there are no hydrodynamic forces on the vehicle (forces due to hydrostatic pressure are not included). This is in contrast with when the vehicle reaches steady state at 15 m ($t=2000$ and on) and the vertical force on the hull is a constant 1000 kN. The ability of the controller with adaptive augmentation to quickly adapt to this additional disturbance and reach steady state is again highlighted by Fig. \ref{fig:sail_planes}. While the non-adaptive controller is also able to reach steady state due to the inclusion of an integral gain in the PID controller, the adaptive controller is able to converge several hundred seconds faster.

\begin{figure}
    \centering
    \includegraphics[width=1\linewidth]{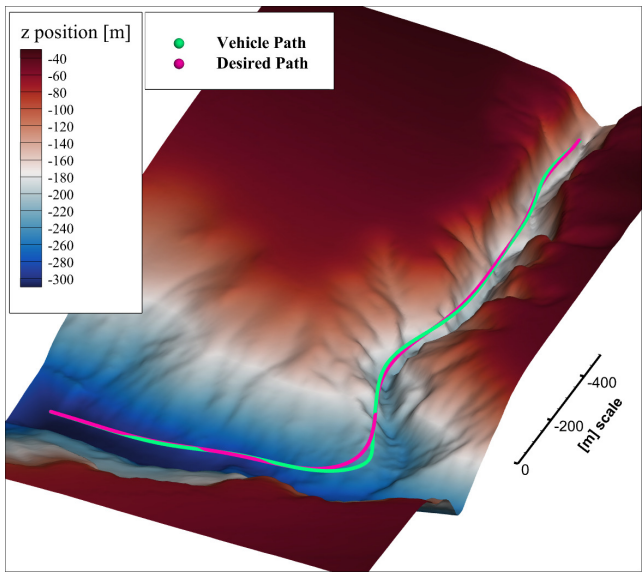}
    \caption{Simulated traversal of the Scripps canyon offshore of San Diego, California by the Joubert BB2}
    \label{fig:canyon_3d}
\end{figure}

Finally, in Fig \ref{fig:canyon_3d}, we demonstrate the capabilities of the controller in a scenario where the vehicle is commanded to traverse a segment of the Scripps canyon near San Diego, California. This is another example of a safety-critical maneuver where guaranteed performance bounds are important to ensure the vehicle will not collide with the canyon walls. The desired path was generated by arbitrarily placing the Bernstein control points such that the path maintained a safe distance from the canyon walls, thus showcasing the ease with which this controller can be used to execute complex maneuvers. Additionally, future work could make use of path-generation techniques such as those used in \cite{cichella2019optimal} to autonomously generate a safe and feasible path.

\begin{figure}
\centering
    \begin{subfigure}{0.5\textwidth}
        \includegraphics[width=\textwidth]{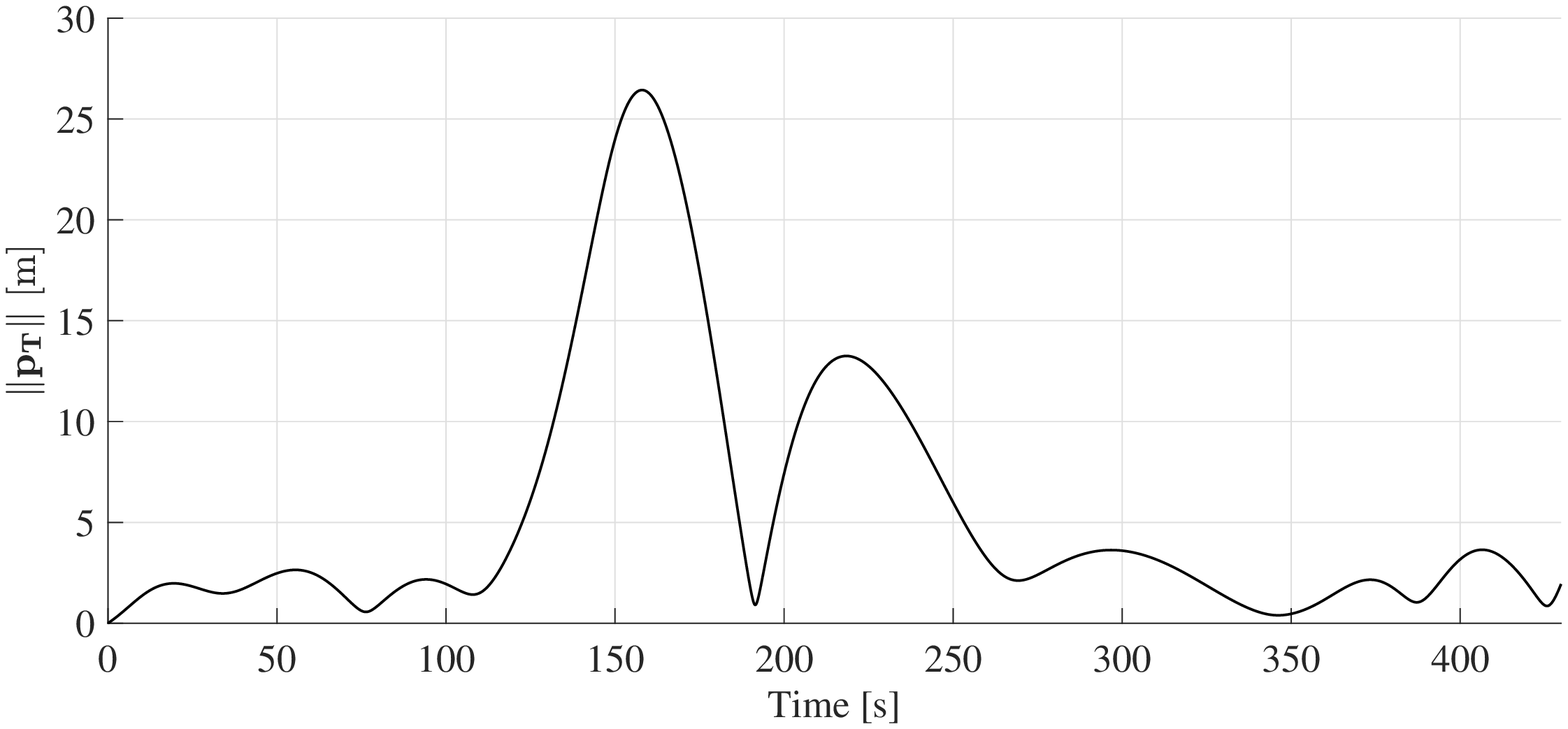}
        \caption{Magnitude of the position error vector $\bm{p_T}$}
        \label{fig:canyon_3d_err}
    \end{subfigure}
    \begin{subfigure}{0.5\textwidth}
        \includegraphics[width=\textwidth]{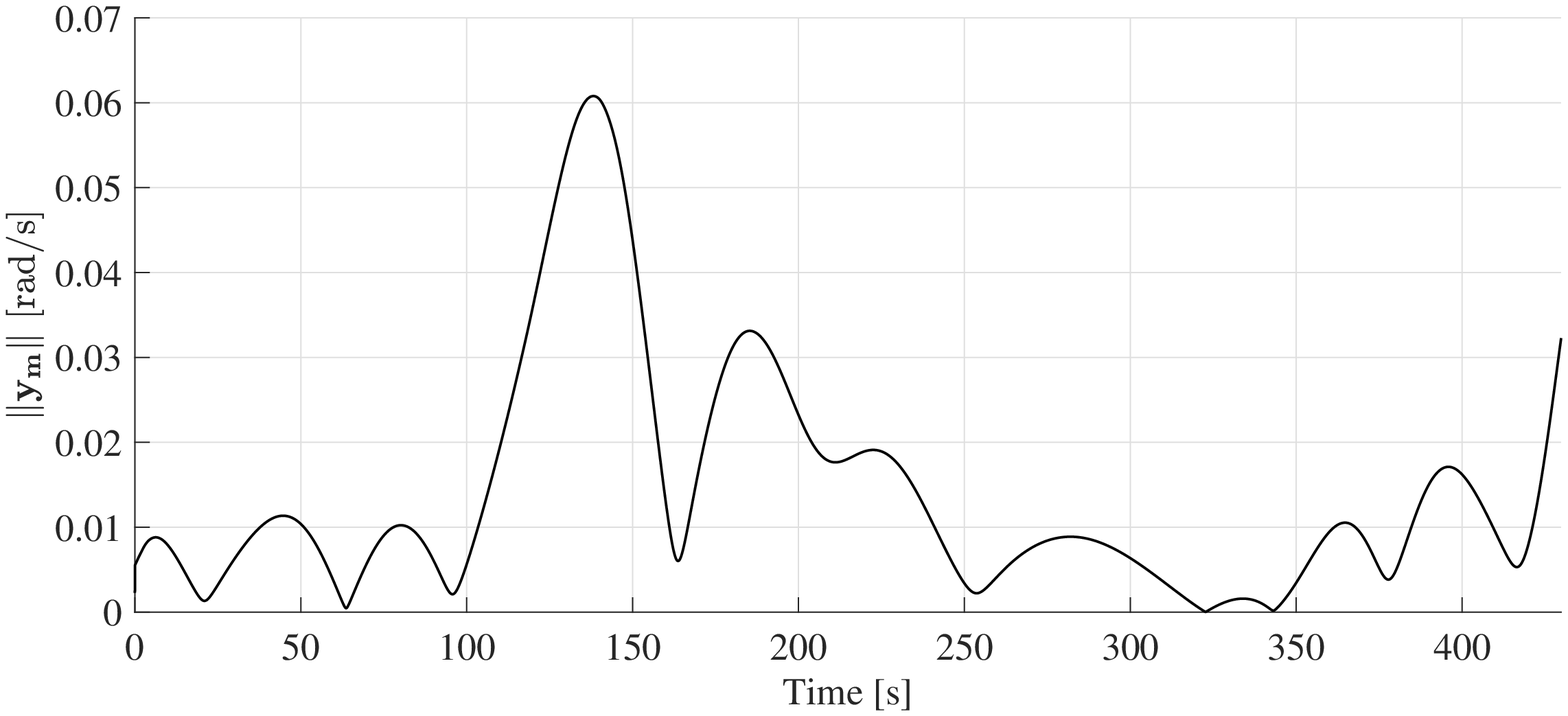}
        \caption{The magnitude of the desired angular rate $\bm{y_m}$}
        \label{fig:canyon_3d_ang_rate}
    \end{subfigure}
    \caption{Position error and desired angular rate through the canyon traversal maneuver}
    \label{fig:canyon_3d_data}
\end{figure}

Additional information for the canyon traversal maneuver is provided in Fig. \ref{fig:canyon_3d_data} which displays the magnitude of the path-following position error vector $\bm{p_T}$ in Fig. \ref{fig:canyon_3d_err} and the components of desired angular rate $\bm{y_m}$ in Fig. \ref{fig:canyon_3d_ang_rate}.
Note that the largest position error corresponds to the sharp turn into the narrower valley. Fig. \ref{fig:canyon_3d_ang_rate} explains this error by showing that the vehicle is being commanded to track relatively large angular rate commands. For context, the vehicle is capable of briefly peaking at 0.05 rad/s in diving maneuvers, but this cannot be maintained for a sustained period of time. Thus, Fig. \ref{fig:canyon_3d_ang_rate} shows that the desired path is not feasible for this vehicle. However, despite this problem, the vehicle is able to reconverge to the path. Additionally, path-generation techniques can be used to avoid this type of issue by considering the vehicle's dynamic constraints when creating the desired path.

\section{Conclusion} \label{sec:conclusion}
This paper has presented a formulation for a multi-layer control scheme for underwater vehicles consisting of a path-following outer-loop controller and an adaptive inner-loop controller. The path-following controller is coordinate-free and determines an angular-rate command to converge to a 3D path. The angular-rate command is passed to the inner-loop controller, which consists of an $\mathcal{L}_1$ augmentation loop that alters the angular-rate command before it is received by an autopilot controller, which are commonly included in commercial vehicles. Performance bounds for both the outer- and inner-loop controllers are derived and presented. Finally, numerical results from a physics-based Simulink model are shown, showcasing the properties of the controller and demonstrating its performance under different conditions. Future work for this project includes implementation of this controller in a lab setting, as well as the addition of design considerations to handle disturbances in the form of waves and changes in dynamics for operation at extreme low speeds.

\bibliographystyle{cas-model2-names}
\bibliography{cas-refs}

\end{document}